\documentclass[doublecol]{epl2} 
\usepackage{graphicx,makeidx,color}
\usepackage{amsmath}                                  
\usepackage{epsfig}
\usepackage{amssymb} 
\usepackage{color}
\title{Coarse-graining complex dynamics:
Continuous Time Random Walks vs.  Record Dynamics.  
}
\author{Paolo Sibani\inst{1}  
 } 

\institute{                    
  \inst{1} FKF,  University of Southern Denmark, Campusvej 55, DK5230, Odense M.
  }
\pacs{nn.mm.xx}{Fluctuation phenomena,  random processes, and Brownian motion}
\abstract{
  Continuous Time Random Walks (CTRW) are widely used to coarse-grain the 
evolution of systems jumping  from 
 a metastable sub-set of  their configuration space, or trap,   to another via  rare intermittent events.
 The multi-scaled behavior typical of complex dynamics is provided by a fat-tailed
 distribution of the waiting time between consecutive jumps.
 We first argue that    CTRW   are inadequate  to describe   macroscopic  relaxation processes 
for three reasons:  macroscopic variables are not  self-averaging,  
 memory effects require  an all-knowing observer, and different mechanisms  
 whereby the jumps affect   macroscopic variables all produce identical long time relaxation  behaviors. 
 Hence, CTRW shed no light on the link between microscopic  and macroscopic dynamics.
We then highlight how a more recent approach,
  Record Dynamics (RD) provides a viable alternative, based on a very different set of  physical ideas:  while CTRW make use of  a renewal process  involving identical traps
of infinite size,  RD   embodies a dynamical 
entrenchment into a hierarchy of traps which are finite in size  and possess different  degrees of meta-stability.
We  show in particular how RD  produces  the stretched exponential, 
power-law and logarithmic  relaxation behaviors ubiquitous  in complex dynamics,  together with the sub-diffusive 
time dependence of the Mean Square Displacement characteristic of  single particles
moving in a complex environment.
}

\begin{document}
 \maketitle

\section{Introduction} 
 Statistical physics is largely about 
 coarse-graining   microscopic descriptions into
 macroscopic ones more closely  related to experiments.
 Thermal relaxation of `glassy' systems is a case in point:
Due to  their  large  number of microscopic configurations from which 
a deterministic (zero temperature)  trajectory  never  escapes,
 configuration space  can be  partitioned  into catchments basins which,   at finite 
temperature,   retain trajectories  for a  lapse  of   time of   finite and random duration. 
We refer to  these basins
as \emph{traps}, to the time spent in them  as  \emph{waiting time}
and to the transitions  between  traps  as \emph{jumps}.
Describing relaxation  in terms of traps and jumps 
greatly reduces the number of variables 
and constitutes   the   first   step of  coarse-graining.
Based on \emph{Continuous Time Random Walks} (CTRW)~\cite{Shlesinger74,Sher75,Barkai00},
 a  well established approach   further assumes  that
 each  jump  brings  the   system  back to the same  situation, i.e. that the
 sequence of  jumps 
  constitutes a   \emph{renewal} process.   
 Using a  fat-tailed distribution for the waiting time, 
 the multi-scaled relaxation  behavior  characteristic of complex systems
   can in many  cases be accounted for. Nevertheless,  a stationary renewal process 
is not a natural choice to 
  describe  the  macroscopic changes occurring in e.g.
 non-stationary relaxation processes. 

 As emphasized in  the much touted \emph{weak ergodicity breaking} scenario~\cite{Bouchaud92,Bel05}
time and ensemble averages  differ
for  renewal processes involving fat-tailed waiting time distributions. 
This property is  closely   related to a well-known mathematical result of Sparre-Andersen~\cite{SparreAndersen53,Feller66b}
by the fact that the number of jumps in the interval $[0,t)$ remains a distributed quantity
in the limit $t\rightarrow \infty$. 
Hence,  in a  CTRW description  macroscopic quantities  have broad distributions
even   in the thermodynamic limit.
A second, related,  issue is related to the system size dependence  of the average and variance of macroscopic 
observables. As we argue, both quantities must scale linearly    with system size, but fail
to do so   in CTRW. Thirdly, the memory mechanism implied by CTRW requires an all-knowing  
observer  and, lastly, the  long-time tail of the waiting time distribution 
can hardly be justified in many applications.
In summary, even though CTRW appear  flexible and eminently applicable,    
their use  to model  complex dynamics is a dubious endeavor.
 We argue below that 
\emph{Record Dynamics}(RD) is a viable   alternative which relies on a completely different
physical picture and which  avoids the problems affecting   CTRW, technically because  
the  jumps are there a Poisson process.

A record in a time series is an entry larger (or smaller)
than all the entries that precede  it. Records have always been a popular topic,
but  a recent  surge of interest in their statistical properties~\cite{Krug07}
seems    motivated by the ongoing  debate on climate change, which
is  accompanied by a number of record breaking events.
 That thermal noise records  have an impact 
 in complex dynamics 
was proposed~\cite{Sibani93a}  
in a model study of Charge Density Waves.
Over the years  the same idea, which we now refer to as Record Dynamics,
has found applications
in  condensed matter physics~\cite{Oliveira05,Sibani06a,Sibani11,Boettcher11}, evolutionary biology~\cite{Anderson04} and the dynamics
of ant societies~\cite{Sibani11a}. 
The term `record' in RD  signals 
that   overcoming   a   record-sized 
dynamical barrier   elicits    a   jump --henceforth in this connection termed  \emph{quake}---
which brings the system from one   trap  
to a new and previously unexplored trap.  RD hence describes a process  of
\emph{entrenchment} into a hierarchy of  traps
indexed   by dynamical barriers  of 
increasing size~\cite{Hoffmann88,Avetisov10}. Focusing on the temporal statistics and
the macroscopic  effects of the quakes,
 RD provides  a coarse-grained description of  glassy dynamics. 
\section{Critique of CTRW}
The probability  $P_{\rm j}(n,t)$ of   $n$ jumps  in the time interval  $[0,t)$
and its first two moments are discussed below, using
the letter $s$  and 
a superimposed tilde  to denote the Laplace    variable and the Laplace transform of a 
function, respectively. 
Central to  the description is the waiting time  probability density (PDF)
$W(t)$.  Whenever its average is  finite, the exponential form
$W(t)=\exp(-t/t_0)/t_0$ is a natural choice and, we stress, a choice  to  which 
our critique does not apply.
To  model complex relaxation a `fat-tailed' PDF  lacking  a finite average
\begin{equation}
W(t) = \frac{\alpha}{t_0} \left(\frac{t}{t_0}\right)^{-\alpha-1}, \quad 0 < \alpha <1, \quad t \ge t_0,
\label{pw_f}
\end{equation}  
is utilized.
Through mathematical steps detailed further below, 
  the average and variance  of the number of jumps occurring in $(0,t)$ are shown, asymptotically
  for large $t$,  to 
  be connected by the equation
   \begin{equation}
  \sigma^2_{{\rm j}} (t) \approx     \mu_{{\rm j}} (t)   + \left( \frac{t}{t_0}\right)^{2\alpha} \left( \frac{1}{\alpha \Gamma(2\alpha)}  - \frac{1}{\Gamma^2(\alpha+1)}\right), 
  \label{variance_n}
  \end{equation}  
   where $\Gamma$ is the gamma function. 
 For $\alpha=1$,    $\sigma^2_{{\rm j}} (t) =   \mu_{{\rm j}} (t) $. Otherwise,
 in the large $t$ limit,   $\sigma^2_{{\rm j}} (t) \propto   \mu^2_{{\rm j}} (t) $
and since  $\sigma_j(t)/\mu_j(t)$ then approaches a constant, the number of jumps
retains a broad distribution  in the same limit.  
As the  same is true for  time   averages
of quantities  subordinated to the jumps but not for the corresponding
ensemble averages,   ergodicity is `weakly' 
broken. 
In contrast,  textbook statistical mechanics teaches us  
  that macroscopic variables are   invariably delta-distributed,  
 including  cases    where broken ergodicity  
 stems from a  broken symmetry.
 To the best of the author's knowledge,  
no experimental evidence has ever contradicted  this result.

 Since a   single  CTRW process cannot consistently 
describe macroscopic  relaxation, let us instead  try  
$N$ independent and simultaneous jumping processes, each supported
in one of  $N$  domains, a 
 situation typical of   spatially extended systems with short-ranged interactions.
The average and the variance
of the number of jumps throughout the system are  in this case both  proportional to $N$, and 
 $\sigma_j(t)/\mu_j(t) \propto N^{-1/2}$ hence   vanishes for  large $N$, 
 taking weakly broken ergodicity along. This sounds reassuring, but, as we 
 shall  see, the memory behavior implied by the description requires an all-knowing observer.

 For any choice of $W(t)$,  renewal equations  for the jump probability $P_{\rm j}(n,t)$,
\begin{eqnarray}
P_{\rm j}(n,t)&= &  \int_0^t P_{\rm j}(n-1,t') W(t-t') dt' \; ;     \quad    n= 1,2 \ldots  \\
P_{\rm j}(0,t)&= &  1 - \int_0^t  W(t') dt',  
\label{renewal}
\end{eqnarray}
are solved in  the $s$  domain by
 \begin{equation} 
  \tilde{P}_{\rm j}(n,s) =   \left( \tilde{W}(s) \right)^n \frac{1 -  \tilde{W}(s) }{s}.
 \label{lap_tr}
\end{equation}
The average  number of jumps, $\mu_{\rm j}(t)  = \sum_{k=0}^\infty k P_{\rm j}(k,t)$
and the auxiliary quantity $\mu_{{\rm j}^2 -{\rm j}}(t)=\sum_{k=0}^\infty (k^2-k) P_{\rm j}(k,t)$
 have then   transforms
 \begin{equation}
 \tilde {\mu}_{\rm j} (s)=    \frac{\tilde{W}(s) }{s (1- \tilde{W}(s))} \quad {\rm and} \quad 
 \tilde{\mu}_{{\rm. j}^2 -{\rm j}} (s)=  \frac{2 }{ s}   \left(\frac{ \tilde{W}(s) }{ 1- \tilde{W}(s)}  \right)^2,
 \label{moments}
 \end{equation} 
 respectively.

To derive Eq.~\eqref{variance_n}, insert
  the  small $s$ expansion 
\begin{equation}
 \tilde {W}(s)= 1- (t_0 s)^\alpha + { \cal O}(s^{\alpha+1})
\label{first_moment_exp}
 \end{equation}
of the Laplace transform of Eq.~\eqref{pw_f} into   Eq.~\eqref{lap_tr}.
 Inverting the outcome  yields 
 \begin{equation}
  \mu_{{\rm j}} (t) \approx  \frac{1}{\alpha \Gamma(\alpha)}\left( \frac{t}{t_0}\right)^\alpha 
  \end{equation}
  and 
\begin{equation}
{\mu}_{{\rm j}^2 -{\rm j}} (t) \approx \frac{1 }{ \alpha\Gamma(2\alpha)}   \left(\frac{ t}{ t_0}  \right)^{2\alpha}.
 \label{sec_moment_time}
 \end{equation} 
The result follows from $\sigma^2_{\rm j}(t) = 
 \mu_{{\rm j}} (t)   +  \mu_{{\rm j^2 -j}} (t)   -  \mu^2_{{\rm j}} (t) $.
  
  If the jumps constitute the true clock 
  of the dynamics, it is natural to describe their effect on
 relaxation as a Markov chain.  The question is then how the 
 properties of the latter affect the relaxation in the time
 domain.
 In general, the propagator of a Markov chain is a linear superposition
  of  exponentially decaying modes, each of the form 
   $b^n$, where $b <1$. The same is true for  averages calculated using
   the propagator. Without loss of generality, we now
   consider the  time dependence $g(b,t)$ corresponding to  a single mode   $b^n$,
 which is   obtained by averaging $n$ over the probability  $P_{\rm j}(n,t)$
 that $n$ jumps occur. In the
 Laplace domain this amounts to 
\begin{equation}
\tilde{g}(b,s) = \sum_{n=0}^\infty \tilde{P_{\rm j}}(n,s) b^{n} = 
 \frac{1 -  \tilde{W}(s)}{s(1-b  \tilde{W}(s) ) }.
\label{LPm}
\end{equation}
If  $W(t)$ has a finite average $t_0$,   
expanding Eq.~\eqref{LPm}  to lowest order, we find  
  that the mode decays exponentially in time, with a time scale $t_0/(1-b)$ 
diverging  as expected for $b \rightarrow 1$.
We also note that  since $g(b,t)$ actually depends on $b$,
 the eigenvalue spectrum of the Markov chain matters  in the time domain.
This  hinges on  the $s$ term in the denominator and  the $t_0 s$ term 
in  the nominator canceling out. The situation radically differs
if  $\tilde{W}(s) = 1 -(t_0 s)^\alpha$ with $0<\alpha <1$. 
To leading order, Eq.~\eqref{LPm}  gives a term proportional to $s^{\alpha-1}$, which in the
time domain     translates into a power-law decay 
 whose   exponent, $-\alpha $, is  independent of $b$.   In other words, 
 the  value of the exponent $\alpha$ is unrelated to the  dynamical effects of  the
 jumps.
 
Consider now  the simple scaling  description 
known as   \emph{pure} or \emph{full aging} behavior,
 which approximately captures some aspects of memory behavior in   glassy dynamics.  
 According to pure aging,  certain macroscopic variables depend on the ratio $\frac{t}{t_{\rm w}}$,
e.g.  in the  thermoremnant
magnetization of spin-glasses~\cite{Sibani11} ,
 $t> t_{\rm w}$ is  the time counted from the 
initial  thermal quench  and 
 $t_{\rm w}$ is the time at which the  external magnetic field is switched off. 
 
Knowing that the   system has remained in the same trap   
up to    time $t_{\rm w}$   
at which   observations commence, the  probability density  for 
exiting  the trap at time  $t > t_{\rm w}$  is  
\begin{equation}
W_R(t_{\rm w},t) = \frac{W(t)}
{
\int_{t_{\rm w}}^\infty
 W(t') dt'} = \frac{\alpha}{t_{\rm w}} \left(\frac{t}{t_{\rm w}}\right)^{-\alpha-1}, 
 \label{rescaled}
\end{equation}
which is identical  to the RD expression \eqref{wait_small_x} obtained below  by 
a different route.
Since  all traps are equivalent in CTRW, the   memory  behavior implied by
 Eq.\eqref{rescaled}  rests on  the observer knowing when a trap is entered.
 This might be experimentally achievable 
 if a single trap describes the whole   system, a possibility  however  already  
discarded as  unphysical. If, however, $N$ independent processes 
unfold at the same time, the observer must track when  every trap
is accessed, a task 
hardly feasible  in experiments.
   
\section{Dynamical hierarchies, records and marginal stability}
Upward rooted binary trees~\cite{Hoffmann88,Sibani91,Sibani93,Sibani94}
whose nodes  and height respectively represent traps  and their  
energy provide a convenient coarse-grained representation of 
energy landscapes with multiple minima.
 In low temperature thermalization, the `bottom' states of lowest energy are
 those mainly occupied, and 
 gaining access to  nodes  not previously visited  entails
crossing an energy barrier of record magnitude. Hence,  diffusion on a hierarchical
structure can be described in terms of RD. In the general case, a   record-sized energy  fluctuation  
does not suffice to elicit  a quake, simply because  there might be  no barrier to cross.
 \emph{Marginal stability}~\cite{Sibani03} further  posits  that 
  the barriers successively crossed   
differ by a minuscule amount. In this  limit
 every  record-sized energy fluctuation leads to the  crossing of a barrier
 and record-sized thermal fluctuations   trigger quakes.  The temporal statistics of the quakes
 occurring between $t_{\rm w}$ and $t >t_{\rm w}$
is  in this limit  a Poisson process,  whose 
average $ \mu_{\rm q}(t_{\rm w},t) \propto (\ln(t) - \ln(t_{\rm w}))$
is  independent of  the temperature~\cite{Sibani93a,Anderson04}.

To  generalize the above results 
to cases where energy barriers differ by  a finite amount, 
consider that, in general,   
    \begin{equation}
\mu_{\rm q}(t_{\rm w},t)  = \int_{t_{\rm w}}^t  r(t') dt',
\label{rate}
\end{equation}
 where $r(t)$ is the quaking rate and where 
 the form $r(t) = a/t$  corresponds to  the logarithmic behavior associated to marginal stability.
 The  generalized form  
  \begin{equation}
r(t) = a t^{x-1}, \quad 0 \le x \le 1,
\label{newrate}
\end{equation} 
\emph{i)} reduces to $a/t$ for  $x= 0$,
\emph{ii)}  produces  time-homogeneous behavior 
for  $x=1$ and,  \emph{iii)} integrated with respect to time,  yields 
\begin{equation}
 \mu_{\rm q}(t_{\rm w},t)  = a \frac{ t^x-t_{\rm w}^x}{x} \stackrel{\rm def}{=} f(t) - f(t_{\rm w}).
\label{newav}
\end{equation} 
Since, as later argued,  a Poisson distribution still applies, a particle happening to reside  in a   trap  at time $t_{\rm w}$
 leaves it at  time   $t>t_{\rm w}$
with  probability  
  \begin{equation}
P_0(t_{\rm w},t) = \exp(-f(t) + f(t_{\rm w})).
\label{P0}
\end{equation}
In terms of   the \emph{lag time} $\Delta  = t - t_{\rm w} \quad (0 \le   \Delta  < \infty)$ commonly used in lieu of $t$, 
the   residence  time $R$  spent   in 
  a trap has PDF 
 \begin{eqnarray}
W_{R}(t_{\rm w},\Delta) &=&
 -\frac{d P_0(t_{\rm w},t_{\rm w}+\Delta)}{d\Delta} \\ \nonumber
  &=& 
 a \exp[
 -\frac{a}{x} \left (
 t_{\rm w} +\Delta)^x -  t_{\rm w}^x 
 \right)
 ]
 \left( t_{\rm w} +\Delta \right)^{-1 +x}.
 \label{wait}
\end{eqnarray}  
 We note in passing that  $\Delta$ is often
denoted by  $\tau$ or by 
$t$ in the literature, both usages unfortunately clashing with our present notation.
  For $x\ll 1$, one obtains
 \begin{equation}
W_{R}(t_{\rm w},\Delta) \approx 
 \frac{a}{t_{\rm w}}  \left (1+\frac{\Delta}{t_{\rm w}} 
 \right)^{-a-1}
 \label{wait_small_x}
\end{equation}  
which is equivalent to the  CTRW expression given by Eq.~\eqref{pw_f}.
Importantly, the time scale parameter which is fixed in CTRW is simply the system age
in RD. Secondly,  Eq.~\eqref{wait} contains a stretched exponential,
and its similarity to $W(t)$ is restricted to 
the limit  $x \rightarrow 0$.
Thirdly, the RD parameter $a$ is positive but  not  \emph{a priori} limited to the unit interval.
For a single  hopping process and in the limit $x \rightarrow 0$,
  if each barrier  record triggers a quake,
$a=1$, otherwise $0 < a <1$.
In  extended systems, where several independent  hopping processes occur
simultaneously, $a$ is proportional to the size of the system, as we later argue.
 
According to   Eq.~\eqref{wait}, the average or characteristic time spent in a trap occupied 
 (but \emph{not} necessarily entered)
at time $t_{\rm w} $ is
   \begin{equation}
 t_0(t_{\rm w}) = a^{-1} \left( \frac{x}{a}
\right)^{\frac{1}{x}-1} \exp(\frac{a}{x} t_{\rm w}^x)\Gamma(\frac{1}{x},\frac{a}{x} t_{\rm w}^x),
\label{average_res}
\end{equation}
where $\Gamma(s,z) = \int_z^\infty \exp(-y) y^{s -1} dy$ is 
the incomplete upper gamma function.
As a check we note that  
 $t_0(t_{\rm w})  = a^{-1} $ for $x=1$ and that in the limit $x\rightarrow 0$
 $t_0(t_{\rm w})  \rightarrow t_{\rm w}$, for $a >1 $. In the same limit and 
 for $a \le 1$, the average is infinite, but $t_{\rm w}$ still provides the  characteristic  time scale
 for the power-law decay implied by Eq.~\eqref{wait_small_x}.

Assume now that an application specific   function   $f$ has been found  such that 
the probability density for the occurrence of a quake is uniform in 
the stretched  observation interval $f(t) - f(t_{\rm w})$. 
Partitioning the interval  into $M$  subintervals, let $p$ be the   probability that a quake falls into any of these
and note that 
the probability for  $n$ quakes occurring  is 
the    binomial $B(p,n,M)$.
In the relevant limit  $p\rightarrow 0$, $M\rightarrow \infty$ and 
$p M\rightarrow  \mu_{\rm q}$, the binomial  tends to
a Poisson distribution, as   claimed.

Using a binary tree to  coarse-grain   an energy landscape~\cite{Hoffmann88},
we  just argued that RD dynamics arises  in  two ways:  in the limit $x\rightarrow 0$,  successive barriers 
increase marginally, records in the impinging noise induce barrier crossings and, on average,
the typical number $n \approx \mu_{\rm q}$  of barriers crossed at time $t$  is proportional to $\ln(t)$.  The Arrhenius relation
$\ln(t) \propto b(n)/T$ where $T$ is the temperature and  $b(n)$ is the  height of the $n'th$ 
barrier 
then implies  $b(n) \propto T n$. 
If marginal stability is relinquished, i.e for  $x>0$,
  we find  $\ln(\mu_{\rm q}) \approx \ln(n(t)) \propto x \ln(t)$
  for $t >>t_{\rm w}$, 
from which we infer   that  the size of the $n'th$ barrier crossed
is $b(n) \propto (T/x) \ln(n)$.

The time dependence  of  a macroscopic 
quantity, say $g$,  is calculated in RD by 
averaging  its   dependence $\tilde{g}(n)$
  over the probability of $n$ quakes 
occurring  in  $(t_{\rm w},t)$, i.e.
\begin{equation}
g(t_{\rm w},t) = e^{-\mu_{\rm q}(t_{\rm w},t)} \sum_{n=0}^\infty \tilde{g}(n) \frac{(\mu_{\rm q}(t_{\rm w},t))^n}{n!},
\label{basic_form}
\end{equation}
where $\mu_q(t_{\rm w},t)$ is given in Eq.~\eqref{newav}.
As a first  example, assume   $\tilde{g}(n) = c(n=0)  b^n$, where $c$ 
expresses the initial condition and where $b<1$.
The stretched exponential behavior ubiquitous in glassy dynamics~\cite{Cipelletti00,Phillips96}
\begin{equation}
g(t_{\rm w},t) = c(t_{\rm w})  e^{-\mu_q(t_{\rm w},t)(1-b)} =    c(t_{\rm w}) e^{-( t^x -t_{\rm w}^x)\frac{a(1-b)}{x}} ,
\label{stretched_exp}
\end{equation}
immediately follows.  If  $g$  is a one-point average, $c(t_{\rm w})= c(t_0)e^{-t_{\rm w}^x\frac{a(1-b)}{x}}$
and  there is in reality only
one time argument $t$.  In contrast, a  two-point  correlation function with $c(t_{\rm w}=1)$
truly depends on two arguments, as well known.

Again using   the lag time $\Delta  = t - t_{\rm w}$  
Eq.~\eqref{stretched_exp} is recast,  for $\Delta/t_{\rm w} \ll 1$ into
\begin{equation}
g(t_{\rm w},{\Delta})  =  c(t_{\rm w}) e^{- \frac{\Delta}{\tau(t_{\rm w})} },
\label{expo_app}
\end{equation}
 an exponential decay with a characteristic time constant
$\tau(t_{\rm w}) = \frac{x t_{\rm w}^{1-x}}{a(1-b)}$.
A  relaxation time increasing with system age 
  is experimentally observed  in colloidal systems~\cite{Cipelletti00,ElMasri09}.
   The age dependence of  the life-time of   
   the exponential  approximation given in Eq.~\eqref{expo_app}   is not usually  discussed,
    but  follows nevertheless 
by   the simple Taylor expansion  given above.
 In the limit $x\rightarrow 0$,
   Eq.~\eqref{stretched_exp} reduces to
  the power-law 
 \begin{equation}
g(t_{\rm w},t) =  c(t_{\rm w}) \left(
\frac{t}{t_{\rm w}} 
\right)^{ -a(1-b)} \approx  c(t_{\rm w}) \exp(-\frac{ a(1-b)}{t_{\rm w}}\Delta),
\label{power_law}
\end{equation} 
where the exponential approximation holds for  $\Delta \ll t_{\rm w}$.
Anticipating a later observation, we now let $\mu_{\rm q}$  be proportional to 
the system size $N$ of a macroscopic system via $a = N \tilde{a}$, where $\tilde{a}$ is a new constant.
Secondly, we treat  $b^n = \exp(\tilde{b}n)$  as  one   
mode of  a relaxation process parameterized by $n$ in lieu of  time.
Of the $N$ eigenvalues in the   spectrum most will 
approach zero as $N \rightarrow \infty$. A glance at  Eq.~\eqref{stretched_exp} shows that only those for which 
$\tilde{b} = {\cal O} (1/N)$ produce 
a macroscopic decay     independent of $N$.
If the decay of  $\tilde{b}$  with $N$  is faster respectively slower  than  $1/N$, the corresponding 
 mode in the time domain   either has a `frozen'    constant value or immediately decays to zero
in the large   $N$ limit. Note  that the stretching exponent $x$ is independent of system size, while 
the exponent $-a(1-b) $ in Eq.~\eqref{power_law} is only  $N$ independent if, as just discussed, $\tilde{b} = {\cal O} (1/N)$.

In summary, simple and general RD arguments lead to dynamical behaviors  common 
in complex systems: stretched exponential relaxation and  power laws with pure aging scaling. The sub-diffusive behavior
of a single particle moving in a complex environment is discussed next.

\section{Subdiffusion}
 Irreversible single particle jumps in complex environments, e.g. binary Lennard-Jones  mixtures in their glassy 
phase~\cite{Vollmayer04}  are indicative of collective configurational re-arrangements. The same is, we believe, 
true for single particle diffusion in  a living cell,  a problem which has recently  been modeled using CTRW~\cite{Jeon11}.
It is difficult to imagine how  a living cell can contain the  traps of infinite, or at least very large, 
 spatial  extension
needed to produce a waiting time distribution with a long-time tail, especially considering that 
the diffusing particle and its  enclosure   have similar  length scales.

Experimental data for dense colloidal system~\cite{Courtland03} re-analyzed in \cite{Boettcher11}
show that single particle Mean Square Displacement (MSD)  grow  logarithmically, a property explained
in  Ref.~\cite{Boettcher11}  using RD.
This   result, which corresponds to the limit $x\rightarrow 0$ in  Eq.~\eqref{variance_vs_time},
  suggests that single particles   in general  probe  
the      local  re-arrangements of their  aging environment.  
This leads to   sub-diffusion formulas 
 rather  similar to their CTRW counterparts. Distinguishing  between the
 two approaches can therefore  be  experimentally challenging,  as it e.g. requires  an analysis of higher moments 
 and/or an explicit investigation of age dependencies via ensemble averages.
To avoid convoluted typography  
the same symbol is used for the moments of the particle position,  irrespective of the method used.
Note however that CTRW  formulas have   one time argument, while RD formulas mostly have two. 

After performing  $n$  independent jumps,  each associated to 
 a random  additive position change $\Delta x_i$, a point particle  is located at 
\begin{equation}
X(n) = \sum_{i=1}^n \Delta x_i.
\end{equation}
Assume for simplicity that the identically distributed $\Delta x_i$  
have  vanishing odd moments and denote their   second and fourth central  moments  
 by $e_2$ and $e_4$, respectively.
 The form of these  moments will depend on e.g. whether the particles move in a potential well, but
 the arguments below do not.
 
 After   $n$ jumps,  the variance of the particle position or, equivalently its  MSD,  is 
$\sigma^2_X(n) = n e_2$. Hence,
\begin{equation}
\sigma^2_X(t) = \mu_{\rm j}(t)  e_2 \quad {\rm and }\quad  \sigma^2_X(t_{\rm w},t) = \mu_{\rm q}(t_{\rm w},t)  e_2 
\label{variance_vs_time}
\end{equation}
for CTRW and for RD, respectively.  Explicitly, using Eq.~\eqref{newav},
we find the sub-diffusive behavior
\begin{equation}
 \sigma^2_X(t_{\rm w},t)=a \frac
 {t^x - t_{\rm w}^x}
 {x} e_2. 
\label{variance_vs_time2}
\end{equation}
Note that if the first jump moment $e_1$ differs from zero a formula of the same type holds 
for the average position.
 Writing for convenience $t_{\rm w} = y $ and  $t=y+ \Delta$, where $\Delta$  is the lag time, 
 and expanding Eq.~\eqref{newav} to first order in $\Delta$, we find 
  \begin{equation}
\mu_{\rm q}(y,y+\Delta) \approx \frac{a}{x^2}\frac{d( y^x)}{dy} \Delta.
\label{small_lag_mu}
\end{equation}
Experimentally, the variance  is  estimated  using the time integral  \begin{equation}
\overline{\sigma^2_X}(t) =  \frac{1}{t_{\rm max} -\Delta}\int_0^{t_{\rm max} -\Delta} \left[ X(y +\Delta) -X(y)  \right]^2 dy,
\label{small_lag_exp}
\end{equation}
where $t_{\rm max}$ is the largest observation time. 
This corresponds to  averaging  $\mu_{\rm q}(y,y+\Delta)$  with respect to
$y$  over the same time span.   
To first order in $\Delta$, the time averaged particle MSD is then 
 \begin{equation}
\overline{\sigma^2_X}(\Delta,t_{\rm max}) \approx \frac{a}{x^2} t_{\rm max}^{x-1} \Delta   \quad {\rm for } \quad \Delta  <  t_{\rm max}
\label{small_lag}
\end{equation}
If, on the other hand, $\Delta \approx t_{\rm max}$, time averaging is of dubious value, and  Eq.~\eqref{variance_vs_time}
directly implies 
 \begin{equation}
\sigma^2_X(\Delta) \approx \frac{a}{x} \Delta^x.
\label{large_lag}
\end{equation}
 Taken together, Eqs.~ \eqref{small_lag} and \eqref{large_lag} describe  a 
 cross-over of the MSD from a  linear  to  a  sub-linear   time dependence, a behavior  
   observed by Jeon et al.~\cite{Jeon11} in their experiments on   lipid granules in an intracellular environment.
These authors claim that their findings `unanimously' point to  CTRW as the mechanism 
behind sub-diffusive behavior, but as we  just  argued   RD offers an alternative explanation. 
 
To  better discriminate between CTRW and RD, consider 
the ratio $B$  between the fourth 
  and the squared  second moment  of $X$.
Given  $n$  jumps, the fourth moment   is
\begin{equation}
  \mu_{X^4}(n) =   n e_4  + (n^2 -n) e_2^2.
  \label{fourth_n}
  \end{equation}  
Correspondingly  in  the time domain  
\begin{equation}
  \mu_{X^4}(t) \approx    \mu_{\rm j}(t)  e_4  + 
  \left( \frac{t}{t_0}\right)^{2\alpha}  \frac{e_2^2}{\alpha \Gamma(2\alpha)}  
  \label{fourth_moment_CTRW}
  \end{equation} 
for CTRW and
  \begin{equation}
  \mu_{X^4}(t_{\rm w},t) =     \mu_{\rm q}(t_{\rm w},t)  e_4  +   (\mu_{\rm q}(t_{\rm w},t))^2 e_2^2
  \label{fourth_moment_RD}
  \end{equation} 
  for RD.
  For CTRW, the   ratio 
  \begin{equation}
B(t)=  \frac{\mu_{X^4}(t)}{(\sigma^2_{X}(t))^2} \approx   \frac{\alpha \Gamma^2(\alpha)}{\Gamma(2\alpha)} + 
 \frac{e_4}{e_2^2} \frac{1}{\mu_n(t)}   \label{general_case_CTRW}
\end{equation}
approaches 
 $ \frac{\alpha \Gamma^2(\alpha)}{\Gamma(2\alpha)}
$
as   $t \rightarrow \infty$.
In the same limit, the  RD expression
\begin{equation}
 B(t_{\rm w},t) =1 + \frac{e_4}{e_2^2} \frac{1}{\mu_{\rm q}(t_{\rm w},t)}
 \label{rRD}
\end{equation} 
approaches unity, independently of the exponent $x$. This  difference offers an  opportunity 
to discriminate between the two  descriptions. Assuming that 
a salient event defining the age of the system can be  identified, a second possibility is to
investigate whether the particle 
MSD has an aging dependence by performing    ensemble averages.  This dependence is present  in RD but
not in CTRW.
\section{Discussion}
 The eminent applicability  of  CTRW  
 conceals  a number of 
 theoretical issues. Firstly,  
  fat-tailed waiting time PDFs for spatially confined processes, such as 
diffusion in cellular environments 
are  in general curtailed by finite size effects.
Secondly,   macroscopic variables modeled with    a single CTRW 
 feature  an unphysical    lack of  self-averaging. 
Finally, since memory in CTRW cannot be rooted in the unchanging  physical properties of the 
traps visited,
it must be rooted  in the observer's  awareness  of the time at which   a trap   is entered. 
This  knowledge is  only  available (in principle) if  traps pertain to  the entire  system,
the possibility already  invalidated by  the  lack of self-averaging.
 
Broadly speaking,  RD  has  the same range of applications as CTRW, but shares none 
of their problems: Residence times have, with a single exception, a finite average which 
increases systematically with system age. This means that, in contrast 
to CTRW,  macroscopic configurations contain  information on
the system' s age, a fact which naturally explains memory behavior  in RD.  
Since quakes  are   a Poisson process, albeit of an  unusual kind, 
subordinated physical  quantities   are always self-averaging.  
Using averages over the number of quakes, RD produces a  wide ranging  
analytical  description of glassy relaxation and  of  single particle diffusion 
in complex environments. 

A  hierarchical configuration space structure
which now seems to find its way into macroeconomics~\cite{Hawkins09},
 was advocated long ago 
by  H. Simon~\cite{Simon62} as a 
defining property of complexity.
Whenever such description  applies,  crossing  record sized  barriers  
 triggers  quakes. 
Conversely, analyzing  the  dynamical effects of record sized perturbations  on the stability of a  system,
a procedure which can in principle  be purely observational, 
provides  important clues   on the configuration space structure.
This line of investigation
has  great potential interest in complex dynamics and can benefit from a recent 
considerable interest in record statistics~\cite{Krug07}.
\bibliographystyle{unsrt}
\bibliography{SD-meld,lap}

\begin{thebibliography}{10}

\bibitem{Shlesinger74}
M.~F. Shlesinger.
\newblock Symptotic solutions of continuous-time random walks.
\newblock {\em J. Stat. Phys.}, 10:421--434, 1974.

\bibitem{Sher75}
H.~Scher and E.~W. Montroll.
\newblock Anomalous transit-time dispersion in amorphous solids.
\newblock {\em Phys. Rev. B}, 12(6):2455--2477, Sep 1975.

\bibitem{Barkai00}
E.~Barkai, R.~Metzler, and J.~Klafter.
\newblock {From continuous time random walks to the fractional Fokker-Planck
  equation}.
\newblock {\em {Phys. Rev. E}}, {61}({1}):{132--138}, {2000}.

\bibitem{Bouchaud92}
J.P.~Bouchaud.
\newblock Weak ergodicity breaking and aging in disordered systems.
\newblock {\em J. Phys. I France}, 2:1705--1713, 1992.

\bibitem{Bel05}
G.~Bel and E.~Barkai.
\newblock Weak ergodicity breaking in the continuous-time random walk.
\newblock {\em Phys. Rev. Lett.}, 94:240602, 2005.

\bibitem{SparreAndersen53}
E. Sparre~Andersen.
\newblock On sums of symmetrically dependent random variables.
\newblock {\em Scandinavian Actuarial Journal}, 1953:123--138, 1953.

\bibitem{Feller66b}
W. Feller.
\newblock {\em An Introduction to Probability Theory and its Applications, vol.
  I}.
\newblock John Wiley, New York London Sidney Toronto, 1966.

\bibitem{Krug07}
J. Krug. \newblock Records in a changing world. \newblock \emph{J. Stat. Mech.} (2007) P07001.
S.N. Majumdar and R.M. Ziff, \newblock Universal Record Statistics of Random Walks and L\'evy Flights
\newblock \emph{Phys. Rev. Lett.} 101, 050601 (2008). \newblock
I. Eliazar and J. Klafter. \newblock Record events in growing populations: Universality, correlation, and aging.
\newblock  \emph{Phys. Rev. E} 80, 061117 (2009).
J. Franke, G. Wergen, and J. Krug, \emph{J. Stat. Mech.} (2010) P10013.
S. Sabhapandit, \emph{EPL}94, 20003 (2011). 
\newblock J.~Franke, G.~Wergen, and J.~Krug.
\newblock Correlations of record events as a test for heavy-tailed
  distributions.
\newblock {\em Phys. Rev. Lett.}, 108:064101, 2012. \newblock 
G.~Wergen and J.~Krug.
\newblock {Record-breaking temperatures reveal a warming climate}.
\newblock {\em {EPL}}, {92}({3}), {2010}.

\bibitem{Sibani93a}
P.~Sibani and P.~B. Littlewood.
\newblock Slow {Dynamics} from {Noise} {A}daptation.
\newblock {\em Phys. Rev. Lett.}, 71:1482--1485, 1993.

\bibitem{Oliveira05}
{L.P. Oliveira, H.J.~Jensen, M.~Nicodemi and P.~Sibani}.
\newblock Record dynamics and the observed temperature plateau in the magnetic
  creep rate of type ii superconductors.
\newblock {\em Phys. Rev. B}, 71:104526, 2005.

\bibitem{Sibani06a}
{P.~Sibani, G.F. Rodriguez and G.G. Kenning}.
\newblock Intermittent quakes and record dynamics in the thermoremanent
  magnetization of a spin-glass.
\newblock {\em Phys. Rev. B}, 74:224407, 2006.

\bibitem{Sibani11}
P.~Sibani and G.G. Kenning.
\newblock Origin of end-of-aging and subaging scaling behavior in glassy
  dynamics.
\newblock {\em Phys. Rev. E}, 81:011108, 2010.

\bibitem{Boettcher11}
S. Boettcher and P. Sibani.
\newblock {Ageing in dense colloids as diffusion in the logarithm of time}.
\newblock {\em {Journal of Physics-Condensed Matter}}, {23}({6}), {FEB 16}
  {2011}.

\bibitem{Anderson04}
{Paul Anderson, Henrik Jeldtoft Jensen, L.P. Oliveira and Paolo Sibani}.
\newblock Evolution in complex systems.
\newblock {\em Complexity}, 10:49--56, 2004.

\bibitem{Sibani11a}
P.~Sibani and S.~Christiansen.
\newblock Non-stationary aging dynamics in ant societies.
\newblock 282:36--40, 2011.

\bibitem{Hoffmann88}
K.H. Hoffmann and P. Sibani.
\newblock Diffusion in hierarchies.
\newblock {\em Phys. Rev. A}, 38:4261--4270, 1988.

\bibitem{Avetisov10}
V.~A. Avetisov and S.~K. Nechaev.
\newblock {Chaotic Hamiltonian systems: Survival probability}.
\newblock {\em Phys. Rev. E}, {81}({4, Part 2}), {2010}.

\bibitem{Sibani03}
P. Sibani and J. Dall.
\newblock {Log-Poisson statistics and pure aging in glassy systems.}
\newblock {\em Europhys. Lett.}, 64:8--14, 2003.

\bibitem{Sibani91}
P.~Sibani and K.H. Hoffmann.
\newblock Relaxation in complex systems : local minima and their exponents.
\newblock {\em Europhys. Lett.}, 16:423--428, 1991.

\bibitem{Sibani93}
P.~Sibani, C.~Sch{\"{o}}n, P.~Salamon, and J.-O. Andersson.
\newblock Emergent hierarchical structures in complex system dynamics.
\newblock {\em Europhys. Lett.}, 22:479--485, 1993.

\bibitem{Sibani94}
P.~Sibani and P.~Schriver.
\newblock Phase-structure and low-temperature dynamics of short range {I}sing
  spin glasses.
\newblock {\em Phys. Rev. B}, 49:6667--6671, 1994.

\bibitem{Cipelletti00}
L.~ Cipelletti, S.~Manley, R.~C. Ball, and D.~A. Weitz.
\newblock Universal aging features in the restructuring of fractal colloidal
  gels.
\newblock {\em Phys. Rev. Lett.}, 84:2275--2278, 2000.

\bibitem{Phillips96}
J~C Phillips.
\newblock Stretched exponential relaxation in molecular and electronic glasses.
\newblock {\em Reports on Progress in Physics}, 59(9):1133, 1996.

\bibitem{ElMasri09}
D.~El~Masri, G.~Brambilla, M.~Pierno, G.~Petekidis, A.~B. Schofield,
  L.~Berthier, and L.~Cipelletti.
\newblock {Dynamic light scattering measurements in the activated regime of
  dense colloidal hard spheres}.
\newblock {\em {Journal of Statistical Mechanics-Theory and Experiment}}, 
  {2009}.
  
\bibitem{Vollmayer04}
K.~Vollmayer-Lee,
 \newblock Single particle jumps in a binary Lennard-Jones system below the glass
   transition.
\newblock {\em J. Chem.  Phys.} 12, (4781), 2004.


\bibitem{Jeon11}
J.-H.Jeon, V. Tejedor, S. Burov, E. Barkai, C.
  Selhuber-Unkel, K. Berg-Sorensen, L. Oddershede, and R. Metzler.
\newblock {In Vivo Anomalous Diffusion and Weak Ergodicity Breaking of Lipid
  Granules}.
\newblock {\em Phys. Rev. Lett.}, {106}({4}), {2011}.

\bibitem{Courtland03}
{R.E. Courtland and E.R. Weeks}.
\newblock Direct visualization of ageing in colloidal glasses.
\newblock {\em J. Phys.:Condens. Matter}, 15:S359--S365, 2003.

\bibitem{Simon62}
H.A. Simon.
\newblock The architecture of complexity.
\newblock {\em Proc. of the American Philosophical Society}, 106:467--482,
  1962.

\bibitem{Hawkins09}
R.J. Hawkins and M. Aoki.
\newblock {Macroeconomic Relaxation: Adjustment Processes of Hierarchical
  Economic Structures}.
\newblock {\em {Economics-The open access open-assessment e-journal}}, {3},
   {2009}.




\end{thebibliography}
\end{document}